\documentclass[%
 aip,
 amsmath,amssymb,
 superscriptaddress,
 reprint,%
]{revtex4-1}

\usepackage{graphicx}
\usepackage{dcolumn}
\usepackage{bm}

\usepackage{float}

\usepackage[utf8]{inputenc}
\usepackage[T1]{fontenc}
\usepackage{mathptmx}
\usepackage{upgreek} 
\usepackage[dvipsnames]{xcolor}
\usepackage{amsmath}
\usepackage{amssymb}
\usepackage{gensymb}

\usepackage{array}

\usepackage{comment}
\usepackage{etoolbox}

\usepackage{soul}
\usepackage{svg}
\usepackage{subfig}

\makeatletter
\def\@email#1#2{%
 \endgroup
 \patchcmd{\titleblock@produce}
  {\frontmatter@RRAPformat}
  {\frontmatter@RRAPformat{\produce@RRAP{*#1\href{mailto:#2}{#2}}}\frontmatter@RRAPformat}
  {}{}
}%
\makeatother
\begin{document}

\preprint{AIP/123-QED}

\title[Interaction of vortex rings generated by two unsynchronized drop impacts]{Interaction of vortex rings generated by two unsynchronized drop impacts}
\author{A. A. L. Huttunen}

\author{G. M. Bessa}%

\author{M. Backholm}
\email{matilda.backholm@aalto.fi.}
\affiliation{Department of Applied Physics, Aalto University, Espoo, Finland}

\date{\today}
             
\begin{abstract}
A liquid drop falling into a pool can create a vortex ring at the right impact conditions. Such drop-formed vortex rings are of importance in nature and technology and the dynamics of rings created by single and synchronised double drop impacts have been extensively studied. In practice, two neighbouring drops rarely impact a liquid surface exactly at the same time, yet the interaction of two unsynchronised vortex rings have not been studied. Here, we have performed experiments with two water drops impacting a water pool at varying time differences $\Delta t$. By using particle image velocimetry, we have quantified the time-evolution of the resulting vortex rings. We find four distinct categories of vortex ring evolution depending on $\Delta t$. At $\Delta t<0.5$ ms, fully symmetric merging of the vortex rings occur. At unsynchronisation larger than this, the merging either becomes asymmetric and incomplete ($0.5 < \Delta t<7$ ms) or non-existent ($\Delta t \geq 7$ ms). At $7 \leq \Delta t<80$ ms, the  creation of the second vortex ring is impaired, whereas at $\Delta t\geq 80$ ms, the first impact no longer affects the formation of the second ring and the two rings evolve without influencing each other. We show that these regimes can be explained by the capillary waves created by the first droplet. Our results demonstrate the importance of the time difference between drop impacts in the creation and subsequent interaction of two adjacent drop-formed vortex rings, which is important for achieving uniform and controlled mixing in high-throughput applications.

\end{abstract}

\maketitle

\section{\label{sec:intro}Introduction}

Vortex rings, that is, vortices with a closed circular axis line, have piqued the interest of researchers for centuries. The same vortex ring properties that led to the popularity of the vortex atom theory \citep{kelvin1867vortex}, namely the mathematically satisfying structure, stability, and omnipresence in nature, continue to make vortex rings widely used test subjects for vortex motion and interaction \citep{shariff1992vortex}. Vortex rings are relatively stable and robust, simple to generate and analyse, and the vortex tube being a closed loop makes it possible to isolate it from the boundaries of the fluid and avoid the end-defects present in other configurations \citep{shariff1992vortex}. Vortex rings and their interactions have been studied extensively due to their significance in many natural phenomena \citep{scott2006vortexstratosphere,barnouin1998lobateness,hoover2017quantifying}, utilization in a wide range of technologies \citep{johnson1984cavitating,akhmetov1980extinguishing,simon2010quantum,olenev2021contactless}, and captivating visual appeal and occurrence in nature and everyday life \citep{lim1992instability,silliman1963william,roskin2009canadian,marten1996ring,mccowan2000bubble}. 

\begin{figure}[t!]
    \centering
    \includegraphics[width=\linewidth]{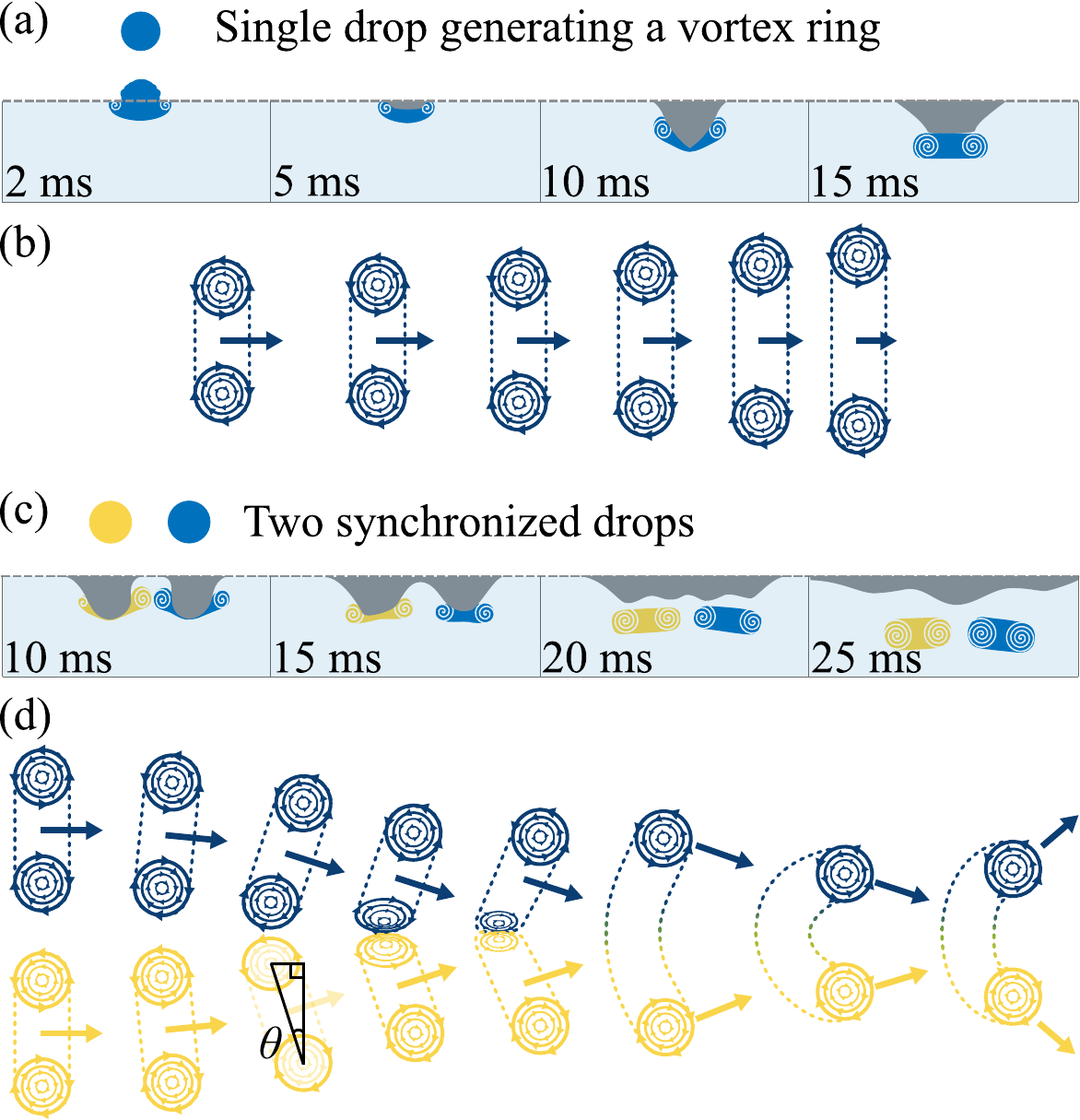}
    \caption{Schematic illustration of \textbf{(a)} a vortex ring generated by a drop impact and \textbf{(b)} its subsequent evolution in a viscous medium. \textbf{(c)} Schematic illustration of the vortex ring generation and crater evolution when two synchronised drops impact a liquid pool, and \textbf{(d)} the topological interaction of the rings coalescing. In \textbf{(a)} and \textbf{(c)}, the pool liquid is illustrated with light blue, the air filled cavity with gray, and the liquid of the two droplets with bright blue and yellow. Drawings inspired by \citep{behera2019generation,oshima1975interaction}.}
    \label{fig:intro_schematic}
\end{figure}

Vortex rings form when a water droplet hits a pool of water (\textbf{figure~\ref{fig:intro_schematic}a}) with the right impact conditions \citep{peck1994three, cresswell1995drop, lee2015origin, behera2019generation}. Drop impacts and coalescence on free surfaces have been an active topic of research due to prevalence in nature \citep{prosperetti1989underwater,michon2017jet} and relevance in applications, such as bioprinting \citep{gupta2025droplet,ng2024jetting}. The impact conditions, such as droplet size, shape, and velocity, affect the motion of the pool surface after impact. A drop impact creates a cavity filled with air as illustrated in \textbf{figure~\ref{fig:intro_schematic}a} with gray. The interaction between the ring and the crater is crucial in determining the vortex ring properties. After detaching from the crater, an undisturbed vortex ring moves straight down due to its self-induced velocity (\textbf{figure~\ref{fig:intro_schematic}b}) and, due to viscous diffusion, the size of the ring increases and the translational velocity decreases \citep{saffman1970velocity, maxworthy1972structure}.

Energy and matter can travel far and fast in vortex rings, which is beneficial in nature, with raindrops enhancing gas exchange in oceans and lakes \citep{saylor2003effect}, as well as in technological applications for contactless fluid mixing and mass transport \citep{lee2015origin, hussain1986coherent, leweke2016dynamics}. Arrays of drops could be used for enhancing the throughput and level of mixing in such applications, but interactions between nearby vortex rings render collisions, thus decreasing their penetration depth and spatial resolution. 
The interaction of two vortex rings moving side by side along parallel axes was studied experimentally by \citeauthor{oshima1975interaction} in liquid\cite{oshima1975interaction} and in air\cite{oshima1977interaction} using dyed water and smoke, respectively. The results of these studies by Oshima and Asaka are the observations of three topological interaction patterns, the time of first collision as a function of initial translational velocities, and the time evolution of the area of the rings viewed head-on. The observed interaction patterns differ from each other by the number of successful collisions. The first collision is illustrated in \textbf{figure 1(d)} with vortex core intersections. Fohl and Turner\cite{fohl1975colliding} showed that the angle at which the rings approach each other determines whether the collision leads to a single merged ring or a second reconnection happens and results in double rings. Vortex ring collision is complex (\textbf{figure~\ref{fig:intro_schematic}d}) and depends on the relative vortex ring sizes, velocities, intensities, and axes. The cores deform, vortex lines stretch, twist, cancel out and reconnect, and the topology changes as explained in detail by \citet{kida1991collision}.

The interaction between two synchronized adjacent drop-formed vortex rings begins on the surface as they are created (\textbf{figure~\ref{fig:intro_schematic}c}). The crater-crater and crater-vortex interactions of two synchronised droplets impacting onto a deep liquid pool have been studied experimentally with dyed droplets by \citet{santini2017experimental} and the crater evolution of multiple synchronised drop impacts has been studied computationally with simulations by \citet{guilizzoni2019synchronized}. Santini \textit{et al.} \cite{santini2017experimental} found that the interaction between the craters created by the two impacts also influences the vortex dynamics, and the translational velocity of the generated vortical structures is lower than in the corresponding single drop impact case. Guilizzoni \textit{et al.} \cite{guilizzoni2019synchronized} concluded that the crater shape does not affect the energy conversion between the kinetic and surface energies significantly, and the maximum depth reached by two merging craters may be larger than in the case of a single impact of a drop with double the volume and mass.


In nature, two neighbouring drops rarely impact on a liquid surface exactly at the same time. Similarly, two close-by vortex rings are more likely to be misaligned than moving perfectly side-by-side. The influence of capillary waves on the dynamics of simultaneous and non-simultaneous double drop impacts on deep pools have been shown by \citet{kirar2022influence} to strongly affect the above-surface coalescence dynamics of the drops. However, to the best of our knowledge, no research has focused on how the time difference, $\Delta t$, between two impacting drops affects the subsequent vortex ring formations or interactions. Here, we have performed experiments with two water drops impacting a water pool at different $\Delta t$ from a falling height rendering clear and strong vortex rings. We have used particle image velocimetry (PIV) to observe and quantify the formation and time-evolution of the two resulting vortex rings. We report four different categories of binary vortex ring evolution depending on $\Delta t$, and based on experimental observations, we suggest that the categories can be explained by how the capillary waves created by the first drop affect the second impact. 

\section{\label{sec:methods}Experimental details}


\subsection{\label{sec:setup}Experimental setup}
\begin{figure}[h]
\centering
\includegraphics[width=\linewidth]{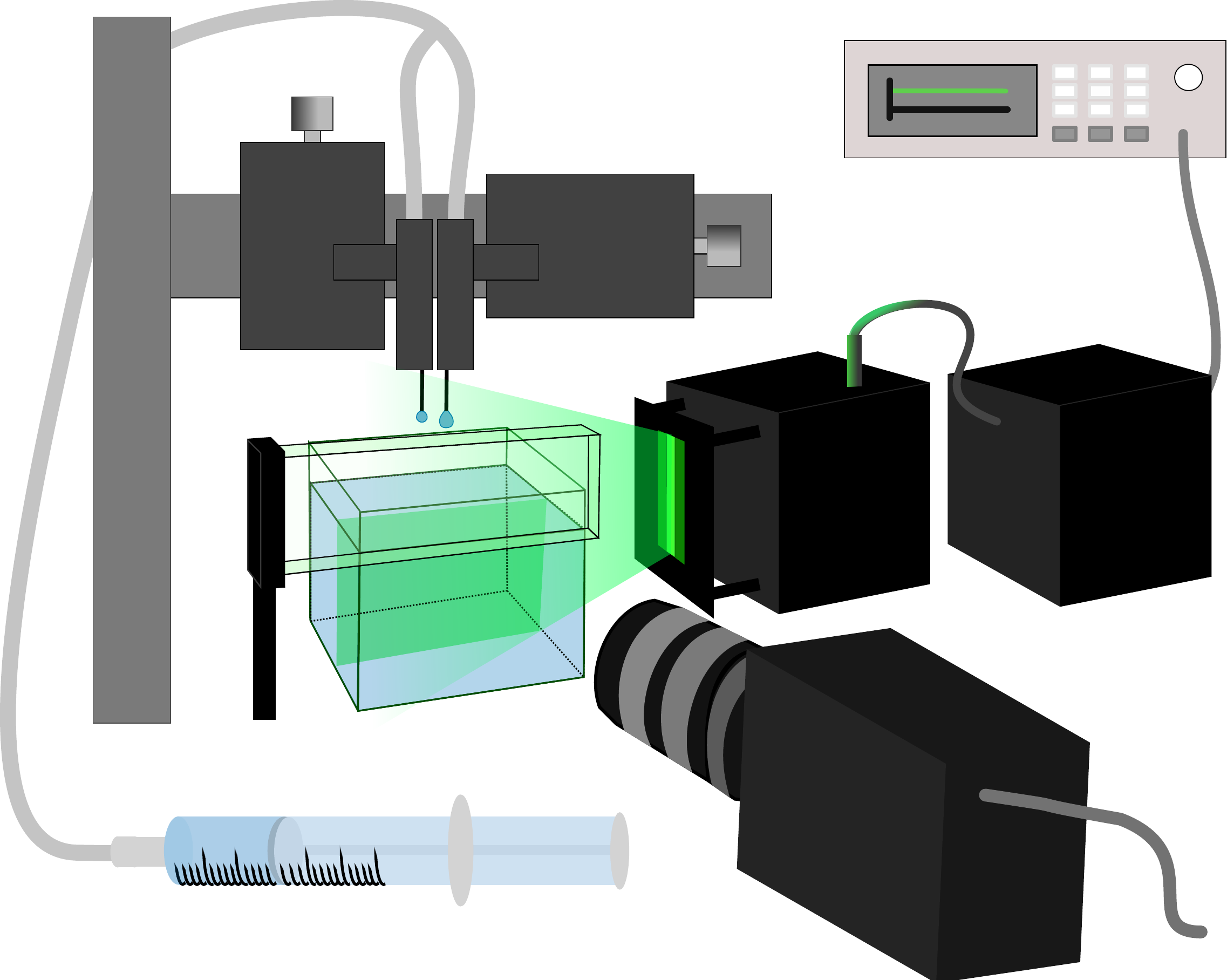}
\caption{\label{fig:setup_schematic} An illustration of the experimental setup consisting of a high-speed camera, LED source, function generator, acrylic block, syringe, two needles and a liquid container.}
\end{figure}

The setup (\textbf{figure~\ref{fig:setup_schematic}}) consists of a clear acrylic container with inner dimensions $7.5 \times 7.5 \times 7.5 \text{ cm}^3$, a high-speed camera (Phantom Miro C211) equipped with a mount adapter (Fotodiox Pro) and objective (Laowa 2:1 Super Macro Lens), a high-power LED source (iLA 5150’s LPS 3), a function generator (Siglent SDG1032X), and two blunt needles (od=1.0 mm, id=0.8 mm) with adjustable spacing attached to one shared 30 ml syringe through plastic tubing. Vortex rings are generated by pushing fluid through the needles so that two equally sized droplets fall into a 4.5 cm deep pool of the same liquid. The liquid contains seeding particles for flow visualization using PIV. The seeding particles were blue polyethylene microspheres (Cospheric, 619-265-12-2-1) with a 45-53 $\upmu$m size distribution and 1.00 g/cc density. The liquid contained deionized water, 0.65 g/l polyethylene microspheres, and 0.13 g/l Polysorbate 20 (Sigma-Aldrich, TWEEN\textregistered 20 P7949). The polysorbate surfactant was required to suspend the insoluble polyethylene microparticles in water. Care was taken to avoid contamination of the fluids, chambers, needles, and tubing as these would affect the drops and vortex rings through variations in the surface tension.

The drops were recorded for 6 mm in free fall above the pool surface to verify the initial conditions in each experiment, and the drop-formed vortex rings were recorded 3 cm deep into the liquid. The difference in the refractive indices of water and air was counteracted by setting a clear acrylic block between the needles and the camera above the surface of water as illustrated in \textbf{figure \ref{fig:setup_schematic}} \citep{reshef2021optic}. Meniscus caused a dark line above the water surface in the recorded frames and obscured the drop at impact. The shape of the impacting drop affects the evolution of the crater which in turn influences the size and translational velocity of the generated vortex ring as observed by \citet{thomson1886v}, \citet{chapman1967formation}, \citet{keedy1967vortex} and \citet{rodriguez1988penetration}, and discussed by \citet{peck1994three}, \citet{cresswell1995drop} and \citet{behera2019generation}, according to whom the most penetrating vortex ring is generated by a prolate shaped drop. For this reason, the shape of the drop at impact was determined by recording the droplets in free fall and the height of fall was adjusted for the impacting drops to be prolate, $S=d_{\mathrm{v}}/d_{\mathrm{h}}\approx1.1$, where $d_{\mathrm{v}}$ and $d_{\mathrm{h}}$ is the vertical and horizontal droplet diameter (\textbf{supplementary figure S1}). The shape in each experiment was confirmed by recording droplets before impact and extrapolating to the moment of impact.

The pool liquid is constantly in motion due to natural convection and remnant movement induced by occasional stirring, previous drop impacts, and vortex rings moving through the liquid. However, this background motion is orders of magnitude slower and more random than the motion caused by the recorded drop impacts and the vortex rings they create. Between each experiment, the pool is left to rest for a few minutes to avoid interactions with capillary waves or flows created by previous drop impacts. The surface level of the pool liquid was kept constant by adding and removing liquid with a separate syringe between experiments. Due to small imperfections in the experimental setup, pushing the liquid out of the syringe leads to the two drops from the two needles detaching almost but not exactly at the same time. The time delay between the two droplets is not controlled with precision, but this imperfection and randomness works in our favor as we want to record varying time differences, which we achieve by repeating the experiment multiple times.

The experiments were recorded with the spatial resolution of 1280 pix $\times$ 720 pix, sample rate 2000 fps, exposure time 390 $\upmu$s, and field of view of 36$\times$20 mm$^2$. The length of one pixel in the area of focus was determined by filming a calibration target (Thorlabs, Grid Distortion Target R2L2S3P2) and the spatial resolution was measured to be 27.89 $\upmu$m/pixel. The images were captured using Phantom Camera Control (PCC 3.8) software and processed in MATLAB (R2024b).

The flow of vortex rings is inherently three-dimensional and especially the collision and merging of two vortex rings leads to complex dynamics that planar PIV cannot fully resolve. Nonetheless, the 2D visualization used here provides valuable insight into the evolving interaction patterns between the rings.

\subsection{\label{subsec:conditions}Drop impact conditions}
The drop impact conditions are reported in \textbf{table~\ref{tab:impact_conditions}}. Experiments were performed with two drops of a similar radius of $R=1.5\pm 0.1$ mm with a separation distance of $L=6.0 \pm 0.1$ mm impacting the pool from a height $h=2.9 \pm 0.1$ cm at different $\Delta t$. These parameters rendered two strong vortex rings at a separation distance that promoted rich interactions, allowing us to isolate the effect of $\Delta t$ on binary vortex-ring dynamics while holding all other physical parameters fixed. Given the time-intensive nature of the experiments and analysis, a broader exploration of the parameter space is beyond the scope of this study. The needle separation distance was measured between the centres of the droplets and kept constant. The normalized nondimensional separation distance in our experiments was $L/d=2.0\pm 0.1$, where $d$ is the mean diameter of the two droplets. The height of fall $h$ was defined as the distance between the tip of the needle and the surface of the liquid pool. The time difference $\Delta t$ between the two drop impacts is defined as $t_2-t_1$, where $t_i$ is the time instance the $i^{\text{th}}$ drop impacts the pool surface. 

\begin{table}
\caption{\label{tab:impact_conditions} Drop impact conditions. The drop diameters of the first ($d_1$) and second ($d_2$) impacting drop and their mean impact velocity and Weber number (We) are listed for all experiments with their respective time difference. The uncertainty in reported diameter values is $\pm 0.1$ mm, impact velocities $\pm 0.01$ m/s, and Weber number $\pm 1$. The bolded rows are the model cases used to illustrate the five different categories (group) throughout this paper.}
\begin{tabular}{c|c|c|c|c|c}
\hline
\hline
        $\Delta t$ (ms) &  Group & $d_1$ (mm) & $d_2$ (mm) & $u$ (m/s) & We 
        \\ \hline
        $\mathbf{<0.5}$& \textbf{(i)} & \textbf{3.0}& \textbf{3.0} & \textbf{0.58} & \textbf{17} \\ 
        \hline
        0.5 & (ii) & 2.9 & 2.9 & 0.58 & 16 \\ 
        0.5 & (ii) & 3.0 & 3.0 & 0.59 & $17$ \\ 
        0.5& (ii) & 3.1 & 3.1 & 0.58 & $17$ \\ 
        2& (ii)  & 3.1 & 3.1 & 0.59 & $17$ \\ 
        \textbf{2.5}& \textbf{(ii)} & \textbf{3.0} & \textbf{3.0}& \textbf{0.58}
        & \textbf{16} \\ 
        2.5& (ii) & $2.9$ & $2.9$ & $0.58$& 16 \\
        2.5& (ii) & $3.0$ & $3.0$ & $0.58$ & 16 \\
        3.5& (ii) & 3.0 & 3.0 & 0.58 & $16$ \\ 
        \hline
        7& (iii) & 3.0 & 3.0 & 0.58 & $16$ \\ 
        13.5& (iii) & 3.0 & 3.0 & 0.58 & $16$ \\ 
        \textbf{17}& \textbf{(iii)} & \textbf{3.1} & \textbf{3.1} & \textbf{0.58} &\textbf{ 17} \\ 
        24.5& (iii) & 3.0 & 3.0 & 0.58 & $16$ \\ 
        40& (iii) & 2.9 & 2.9 & 0.59 & 16 \\ 
        56.5& (iii) & 3.0 & 3.1 & 0.58 & 17 \\ 
        \hline
        \textbf{80}& \textbf{(iv)}  & \textbf{3.1} & \textbf{3.1} & \textbf{0.58} & \textbf{17} \\ 
        100& (iv)  & 3.1 & 3.1 & 0.58 & 17 \\ 
        169& (iv)  & 3.1 & 3.1 & 0.58 & 17 \\ 
        \hline
        \textbf{"\pmb{$\infty$}"} & \textbf{Single} & \textbf{2.9}& \textbf{--} & \textbf{0.58} & \textbf{16} \\ 
        "$\infty$"& Single & 3.0& --  & 0.59 & 17 \\ 
        "$\infty$"& Single & 2.9 & -- & 0.59 & 16 \\
        \hline
\hline
    \end{tabular}
\end{table}

\begin{figure*}[!t]
\includegraphics[width=\linewidth]{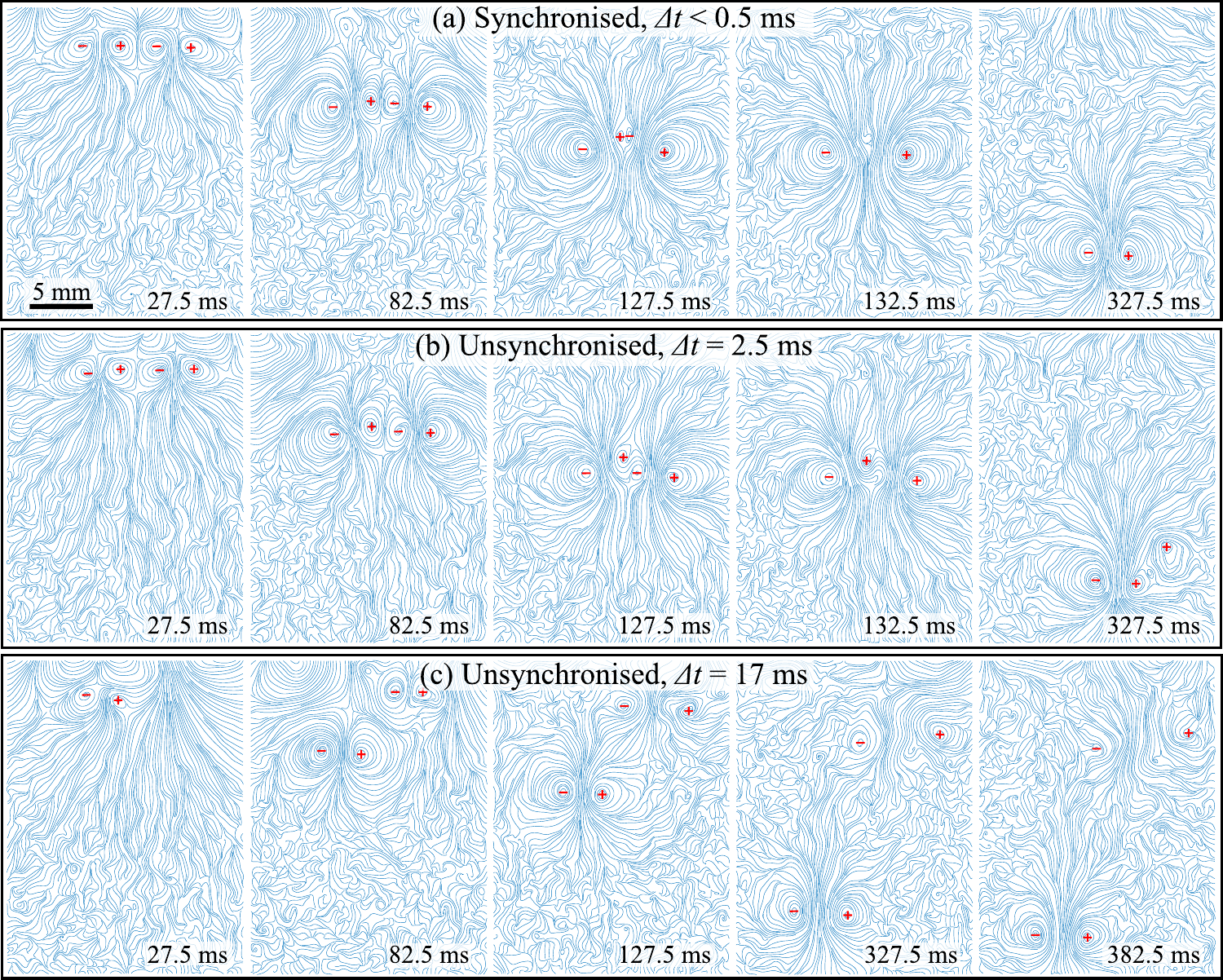}
\caption{\label{fig:streamlines} Streamlines at five different time instances in three experiments performed with \textbf{(a)} synchronised and \textbf{(b-c)} unsynchronised drops. The vortex core centres are indicated with red + and - signs indicating the direction of circulation. The time instances are chosen to show the rings after detaching from the craters, before and after merging (when applicable), and near the end of the recording.}
\end{figure*}

The volume of a drop is estimated by fitting an ellipse on the drop in all of the frames in which the drop is visible during its free fall and calculating the volume of an ellipsoid with corresponding maximum and minimum diameters. The equivalent radius and diameter of a spherical drop is calculated from the mean volume, and the maximum and minimum diameters are used to determine the shape of the drop at each frame (\textbf{supplementary figure S1}).The reported drop diameters ($d_1$ and $d_2$) in \textbf{table \ref{tab:impact_conditions}} were obtained by taking the mean of the diameters from over ten frames of the droplet in free fall. The standard deviation in each case was $<0.01$ mm. The centre of the fitted ellipse is recorded to estimate the impact velocity reported in table \textbf{table \ref{tab:impact_conditions}} by fitting a line through the location data across the analysed frames of free fall. The standard error of each linear fit was at maximum $<0.01$ m/s. The Weber number is calculated as $\text{We}=\rho u^2d/\gamma$, where $\rho$ is the density of the liquid, $u$ the impact velocity, and $\gamma$ the surface tension. The surface tension of the liquid was measured using the pendant drop method \citep{andreas2002boundary,BERRY2015226,goy2017surface} and determined to be $\gamma=0.062\pm0.008$ N/m (\textbf{supplementary figure S2}).

\subsection{\label{subsec:analysis}Data analysis}
The recorded frames were analysed in MATLAB using PIVlab toolbox, a free open-source PIV software \citep{stamhuis2014pivlab,Thielicke_2021}. In preprocessing, the region of interest was set to the entirety of the field of view under the liquid surface, and the craters were masked in all relevant frames. Multipass FFT window deformation PIV algorithm was used with an interrogation window area of 48 pixels in Pass 1 and 24 pixels in Pass 2 with a 50\% step.




\begin{figure*}
    \centering
    \includegraphics[width=\linewidth]{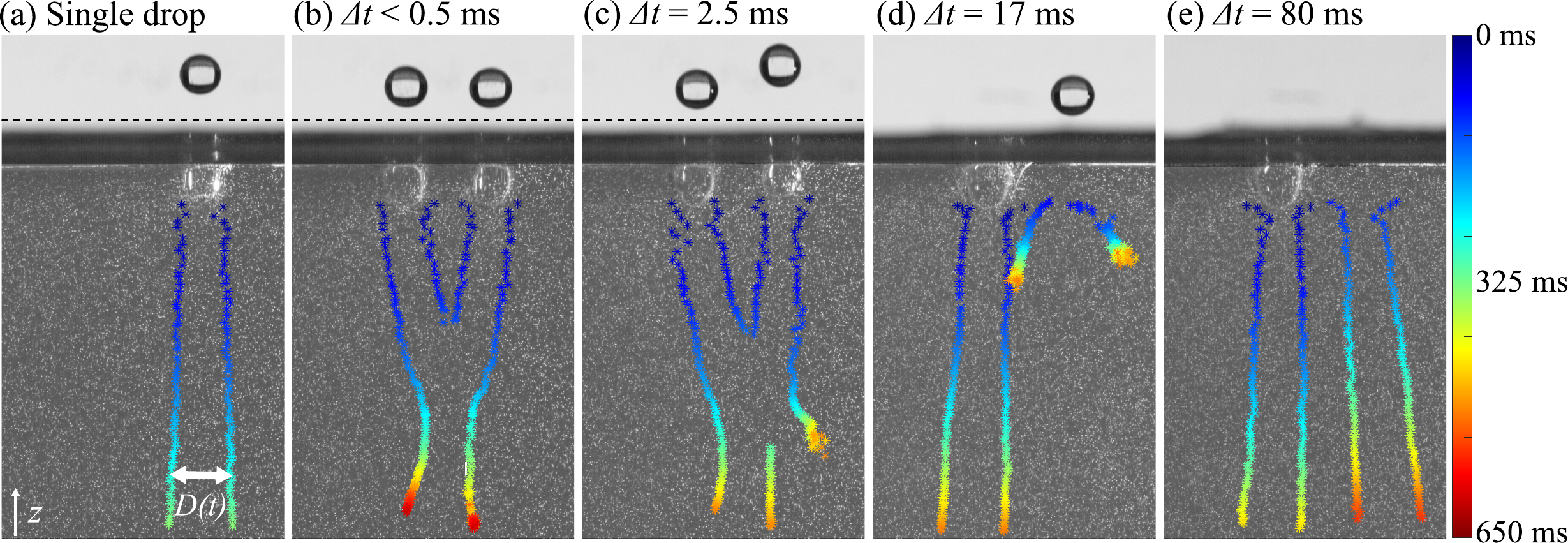}
    \caption{Trajectories of vortex ring centres created by single, synchronised, and unsynchronised drop impacts. The colour of each point corresponds to time after first impact. (a) Single drop impact leading to an undisturbed vortex ring. (b) Synchronised drop impacts leading to symmetrically merging vortex rings. (c) Minorly unsynchronised impacts that lead to asymmetrically merging vortex rings. (d) Critically unsynchronised impacts that lead to a successful vortex ring and a heavily impaired ring. (e) Majorly unsynchronised impacts leading to two vortex rings.}
    \label{fig:spatial_traj}
\end{figure*}

\section{\label{sec:results}Results}

\subsection{\label{subsec:traject}Trajectory}

Vortex core centres were located by plotting streamlines based on the velocity fields, and manually clicking the centres of the smallest circular loops as illustrated in \textbf{figure~\ref{fig:streamlines}} with red markers. In \textbf{figure~\ref{fig:streamlines}a}, the streamlines in a synchronised double drop experiment are shown at five points in time. The collision of the two vortex rings is symmetric, and after merging, only two cores are visible in the streamlines. The observed symmetric merging aligns with previous reports in literature. The initial distance between the rings is small enough to mutually induce acceleration opposite to the self-induced translational velocity \citep{oshima1977interaction}. This causes the rings to tilt and move towards each other, deforming the cores. At the point of contact, vorticity is annihilated due to the opposite direction of circulation. The vortex filaments reconnect and the two rings merge to form a single distorted ring as observed by \citet{oshima1975interaction}. According to \citet{kida1991collision}, the reconnection is never complete.

\citet{kida1991collision} described a single collision by dividing it into three phases: (I) inviscid advection bringing antiparallel parts of the vortex rings together leading to vortex line stretching and core deformation, (II) annihilation and bridging of antiparallel vortex lines, and (III) developed bridges reversing the curvature of the unreconnected anti-parallel vortex lines causing them to move apart and remain as small threads. The small leftover threads decay in interaction with the main ring. If there is enough energy, the outer cores continue to approach each other and subsequent collision and cross-linking of the vortex filaments can split the distorted ring into two smaller rings composed of equal parts of the originals \citep{oshima1975interaction}. If there is not enough energy for a second collision and reconnection, the cores start to move away from each other, the curved shape starts to relax, and the shape of the distorted ring oscillates. We only observe a single collision leading to a single merged ring in \textbf{figure~\ref{fig:streamlines}a}.

In \textbf{figure~\ref{fig:streamlines}b}, the streamlines are shown at the same points in time as in \textbf{figure~\ref{fig:streamlines}a} but in an unsynchronised double drop experiment with $\Delta t=2.5$ ms. After the asymmetric collision, there are three visible vortex cores remaining by the end of recording. Even though the two middle cores disappear from the streamlines for a while after the collision, eventually the right side core of the left ring reappears between the outer cores and continues moving down next to the outer left core. The reconnection between the two colliding rings is clearly far more incomplete than in \textbf{figure~\ref{fig:streamlines}a}. In \textbf{figure~\ref{fig:streamlines}c}, the streamlines are shown for an unsynchronised double drop experiment with $\Delta t=17$ ms. The second ring seems to be severely impaired from the beginning and unable to penetrate deep into the pool like the first ring.

\begin{figure*}[!t]
    \centering
    \includegraphics[width=0.85\linewidth]{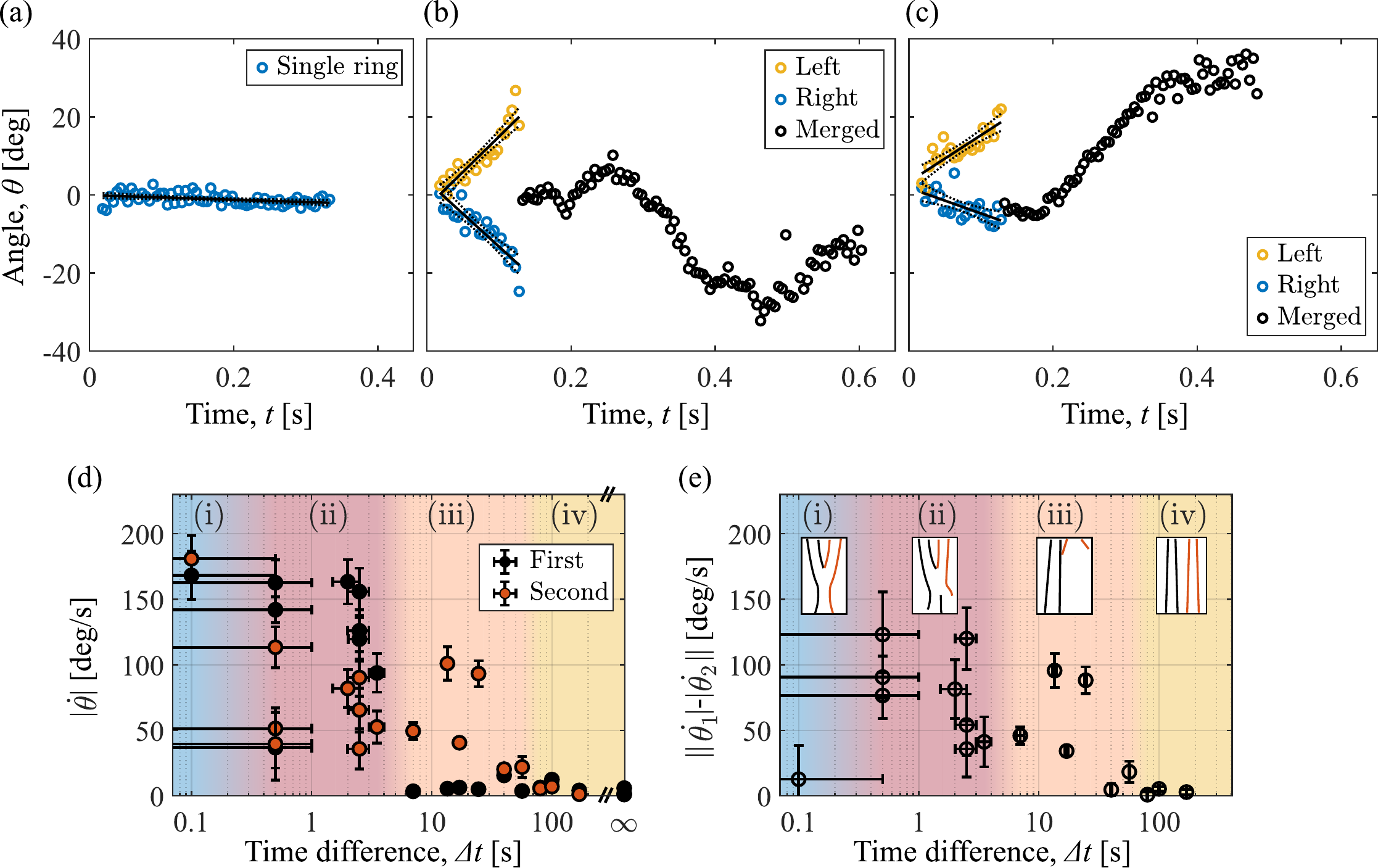}
    \caption{Vortex ring angle as a function of time for \textbf{(a)} a single undisturbed ring, \textbf{(b)} a pair of synchronised rings ($\Delta t<0.5$ ms), and \textbf{(c)} a pair of unsynchronised rings ($\Delta t=2.5$ ms). The left drop impacted on the surface before the right drop. \textbf{(d)} Angle change rates before merging as a function of time difference between drop impacts. \textbf{(e)} Difference in angle change rates. The four time difference regimes based on the interaction patterns of the vortex rings are indicated with shaded regions: (i) symmetric merging, (ii) asymmetric merging, (iii) second ring significantly impaired, and (iv) two strong vortex rings that do not collide. The error bars in (d) are the standard errors obtained from the linear fits and in (e) the error propagation of standard errors.}
    \label{fig:angle}
\end{figure*}

Examples of the trajectories of visible vortex cores from a single drop experiment, a synchronised double drop experiment, and three unsynchronised double drop experiments are presented in \textbf{figure~\ref{fig:spatial_traj}} (see also the corresponding \textbf{supplementary movies M1-M5}). The colours of the data points correspond to the time elapsed since the first drop impact. Four different double-ring categories can be identified based on the trajectories: (i) symmetric merging, (ii) asymmetric merging, (iii) second ring impaired, and (iv) two independent rings. The observed behaviour depends on the time difference between the drop impacts. These different regimes will be discussed in detail below.

The vortex rings generated by single drops (\textbf{figure~\ref{fig:spatial_traj}a}) follow a straight path down and the ring diameter $D$ increases due to viscous diffusion, in agreement with \citet{tinaikar2018understanding} and \citet{behera2021viscous}. The two vortex rings generated by two fully synchronised drop impacts ($\Delta t< 0.5$ ms) approach each other (\textbf{figure~\ref{fig:spatial_traj}b}), and the two cores that are pushed together eventually disappear. The outer two cores then continue to approach each other until they reach a minimum distance similar to that of the initial distance between the two colliding cores. Finally, the distance between the cores starts to slowly grow until the ring is no longer in the field of view. The observed collision and merger fit the interactions observed by \citet{oshima1975interaction}, even though, to the best of our knowledge, the trajectories have not been previously reported. The fact that the cores stop approaching each other indicates that there is not enough energy for a second collision that would break the merged ring in two. Symmetric merging trajectories were also observed in 3 out of 8 experiments with time differences of $0.5\leq\Delta t\leq 2.5$ ms, but further analysis of their tilting angles (Sect.~\ref{subsec:angle}) reveals asymmetric behaviour.

Small time differences of $0.5\leq\Delta t<7$ ms between the two impacts lead to asymmetric trajectories in most experiments. This asymmetric regime (ii) is illustrated with an example trajectory from an unsynchronised experiment with $\Delta t = 2.5$ ms as shown in \textbf{figure~\ref{fig:spatial_traj}c}. The first ring moves more towards the second ring before they collide. The colliding cores disappear initially from the streamlines, and the remaining outer cores approach each other until the outer core of the second ring starts to slow down, as if it is losing its energy, and move away from the collision plane. As the outer core moves away, another core with the same sign of vorticity seems to re-appear between the two outer cores. The re-appeared core moves down in sync with the remaining visible core of the first ring. It seems that the first ring survives the collision and the second ring gets destroyed, but without knowledge of the complete 3D flow, this interpretation cannot be confirmed.

\begin{figure*}[!t]
    \centering
    \includegraphics[width=0.8\linewidth]{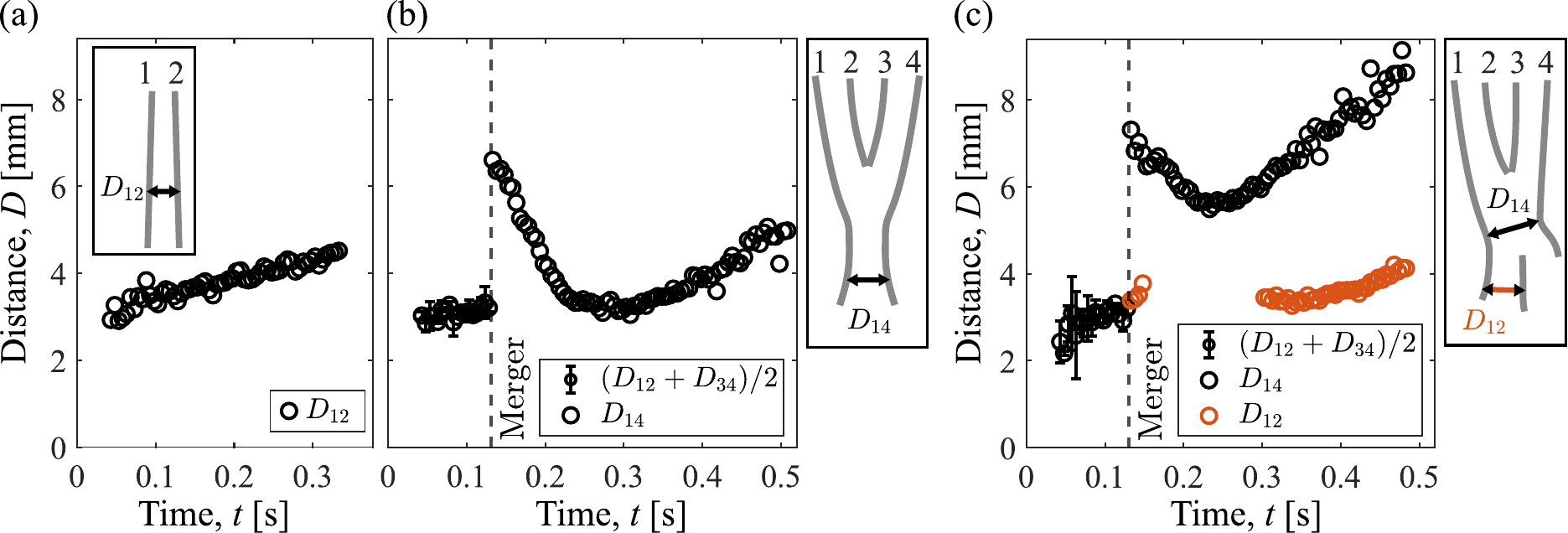} 
    \caption{The distance between the centres of the two visible core intersections in \textbf{(a)} a single undisturbed ring, \textbf{(b)} a pair of synchronised rings ($\Delta t<0.5$ ms), and \textbf{(c)} a pair of unsynchronised rings ($\Delta t=2.5$ ms). Before the merging of rings, the distance between the cores matches the ring diameters, but after the collision, the ring is distorted and the distance between the visible cores is smaller than the actual diameter of the ring. The errorbars in \textbf{(b)} and \textbf{(c)} for $D$ before merging is the deviation of $D_{12}$ and $D_{34}$ from their average.}%
    \label{fig:D}%
\end{figure*}

When the time difference between the impacts is $7\leq\Delta t<80$ ms, the surface interactions dominate the observed behaviour. This time-difference regime covers a wide range of interactions between capillary waves, craters, and vortex rings which affect the generation of the second vortex ring (see \textbf{supplementary movies M4} and \textbf{M6-M8}). In all cases the second ring is significantly impaired and the ring is not able to penetrate deep into the liquid. The experiment with $\Delta t=17$ ms illustrates one of these cases in \textbf{figure~\ref{fig:spatial_traj}d}. By the time of the second impact, the first crater has reached the impact area of the second drop. The second drop impacts on a non-horizontal dynamic surface, and the second impact does not lead to a clean strong vortex ring able to penetrate deep into the liquid. Furthermore, the first ring is not completely unaffected by the second: it moves down slower than an undisturbed ring and the trajectory is not perfectly vertical. The first ring seems to be gently pushed away from the second.

When the time difference is $\Delta t\geq 80$ ms, the surface perturbations caused by the first impact no longer affect the second impact, and the generated vortex rings are far enough from each other not to approach each other and collide. An example is shown in \textbf{figure~\ref{fig:spatial_traj}e}. Increasing the time difference between the impacts eventually leads to effectively single undisturbed vortex rings.

\subsection{\label{subsec:angle}Angle}
In \textbf{figure~\ref{fig:angle}a}, the angle evolution ($\theta$ defined in \textbf{figure~\ref{fig:intro_schematic}d}) of a vortex ring generated by a single drop impact is plotted as a function of time. The zero angle of $\theta =0 ^\circ$ corresponds to a fully horizontal ring orientation. The constant zero angle seen in \textbf{figure~\ref{fig:angle}a} confirms that the single ring is undisturbed and moves straight down the entire time it is recorded.

Before two parallel vortex rings merge, they tilt as illustrated in \textbf{figure~\ref{fig:intro_schematic}d}. According to previous studies \citep{oshima1975interaction}, the ring shape is deformed after merging, but the two visible vortex core cross-sections continue moving faster than the rest of the ring and approaching each other. The vortex ring angle data from the synchronised experiment is plotted in \textbf{figure~\ref{fig:angle}b} as a function of time. Time $t=0$ corresponds to the moment the first drop comes into contact with the surface of the pool liquid. Before merging at $t=0.13$ s, the left ring tilts toward the right ring at the same rate as the right toward the left ring. Right after the two rings have merged, the angle between the remaining cores is close to zero. The angle oscillates around zero after merging. 

In \textbf{figure~\ref{fig:angle}c}, the angle evolution in an experiment with $\Delta t=2.5$ ms is plotted as a function of time. Before merging, the angle of the first (left) vortex ring grows at a higher rate than the angle of the second (right) ring which matches the asymmetric trajectories observed in \textbf{figure~\ref{fig:spatial_traj}c}, where the first ring moves more toward the second ring than vice versa. In \textbf{figure~\ref{fig:spatial_traj}c}, the remaining core of the second ring is slightly ahead of the remaining core of the first ring immediately after the collision, but this order changes quickly. The core of the first ring continues to move down significantly faster than the decaying core of the second ring as the right core of the first ring re-appears between and ahead of them. This tilts the merged ring significantly to the side of the second drop impact (right), which is seen in the angle evolution after merging in \textbf{figure~\ref{fig:angle}c}.

\begin{figure*}[!t]
    \centering
    \includegraphics[width=0.85\linewidth]{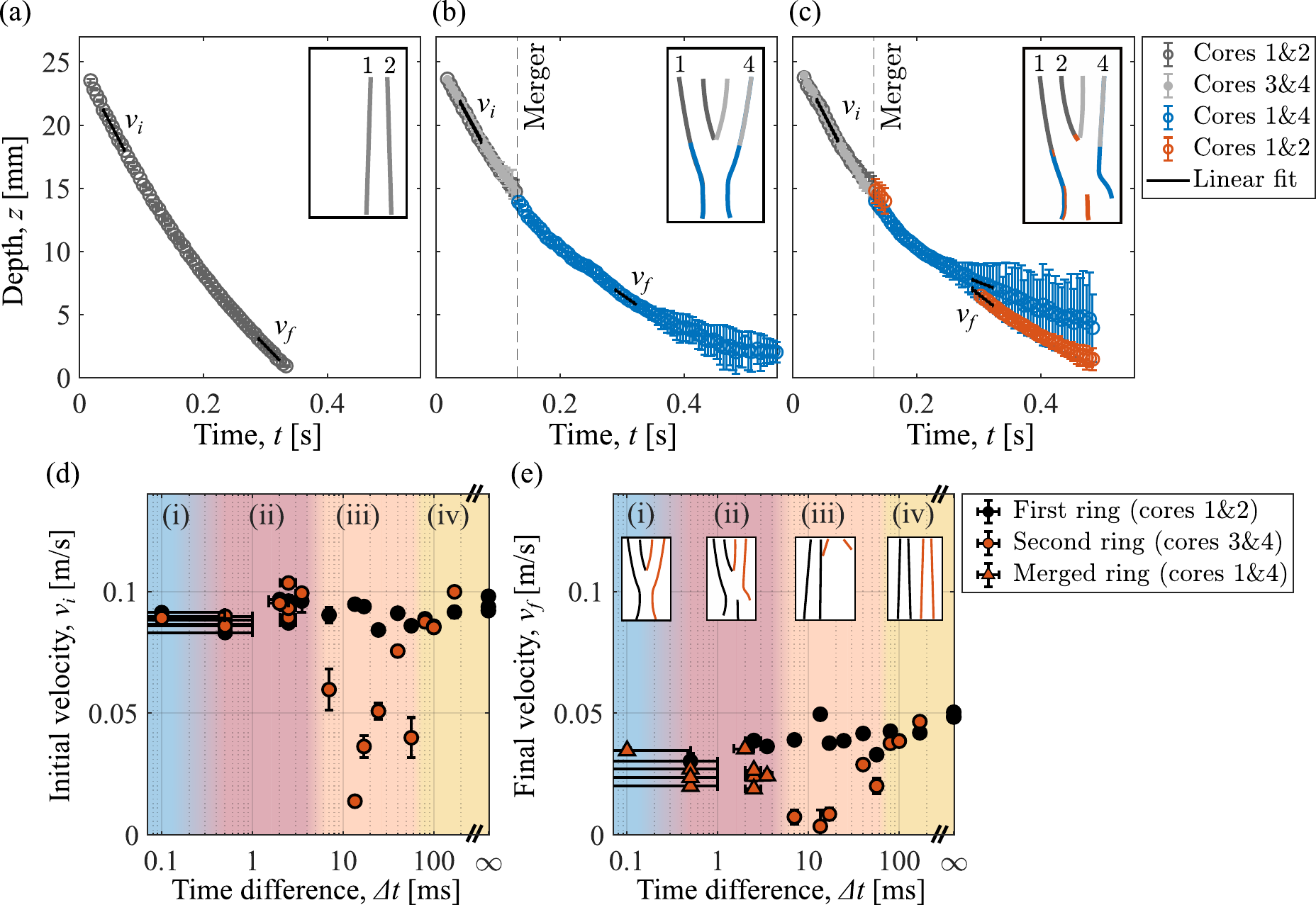}
    \caption{The average depth of vortex core pairs in \textbf{(a)} single, \textbf{(b)} synchronised ($\Delta t<0.5$ ms) double, and \textbf{(c)} unsynchronised ($\Delta t=2.5$ ms) double drop experiments. The \textbf{(d)} initial velocity and \textbf{(e)} final velocity as a function of the time difference between drop impacts. The error bars in \textbf{(a)}-\textbf{(c)} denote the deviation from the plotted average depth of the vortex core pairs. The error bars in \textbf{(d)} and \textbf{(e)} denote the temporal resolution (0.5 ms) and the standard error obtained from the linear fits.}%
    \label{fig:vels}%
\end{figure*}

The angle change rates of each vortex ring in the experiments is plotted in \textbf{figure~\ref{fig:angle}d} as a function of time difference between drop impacts. There is a clear trend of the angle change rate decreasing as the time difference increases. Four time difference regimes can be identified from the graph: (i) in the symmetric merging regime the angle change rates are the highest and symmetric, (ii) in asymmetric merging the first ring has a high angle change rate and the second a lower non-zero change rate, (iii) in the critical regime the first ring moves almost straight down, but the second tilts, and (iv) both rings move straight down. In order to fit the synchronised case in the logarithmic graph, its time difference was set to $\Delta t=0.1$ ms. Our temporal resolution was 0.5 ms due to the sample rate of 2000 fps, so the synchronised case may have a time difference of $0\leq\Delta t< 0.5$ ms. 

The differences between the angle change rate magnitudes of vortex ring pairs in all double drop experiments are shown in \textbf{figure~\ref{fig:angle}e}. In the synchronised case, the orientations of the rings change symmetrically and the difference between the angle change rates is near zero. In the unsynchronised cases where the time difference between the drop impacts is $0.5\leq\Delta t<7$ ms, the second ring tilts less than the first ring before merging. Therefore, the difference between angle change rates is significant. In the $7\leq\Delta t<80$ ms cases where the second ring is heavily impaired, the first ring travels relatively straight down, but the orientation of the second ring usually changes due to its instability. When the time difference between the two impacts is  $\Delta t\geq 80$ ms, the rings behave almost as single undisturbed vortex rings and travel straight down.

\subsection{\label{subsec:dist}Distance between cores}
The temporal evolution of the distance $D$ between the visible vortex core centres is presented in \textbf{figure~\ref{fig:D}} for the same experimental cases as in \textbf{figure~\ref{fig:angle}}. After the vortex ring has detached from the crater and reached its minimum radius, the radius increases with time due to viscosity and the entrainment of pool liquid in the vortex ring, in agreement with previous studies \citep{tinaikar2018understanding}. This is most visible in the single drop experiment shown in \textbf{figure~\ref{fig:D}a}, since it has the longest time to evolve without disturbance.

After two synchronised rings merge in \textbf{figure~\ref{fig:D}b}, the distance between the remaining two visible core centres is approximately the sum of the diameters of the individual rings. The merged ring is deformed and thus the distance between the visible cores does not correspond to the actual size of the ring. The distance between the cores of the merged ring decreases until it reaches a minimum value similar to the diameters of the individual rings before merging, and starts to increase again.

The diameters of vortex rings formed by unsynchronised drop impacts in \textbf{figure~\ref{fig:D}c} differ more than in the synchronised case. After merging, the distance of the cores decreases slower than in the synchronised case in \textbf{figure~\ref{fig:D}b}, and also starts to increase earlier and faster as the remaining core of the second ring decays and the other core of the first ring re-appears. The distance between the two cores of the first ring ($D_{12}$) is similar to a single undisturbed ring diameter. The time derivative of the distance $D_{12}$ is also significantly smaller than of distance $D_{14}$.

\begin{figure*}[!t]
    \centering
    \includegraphics[width=0.85\linewidth]{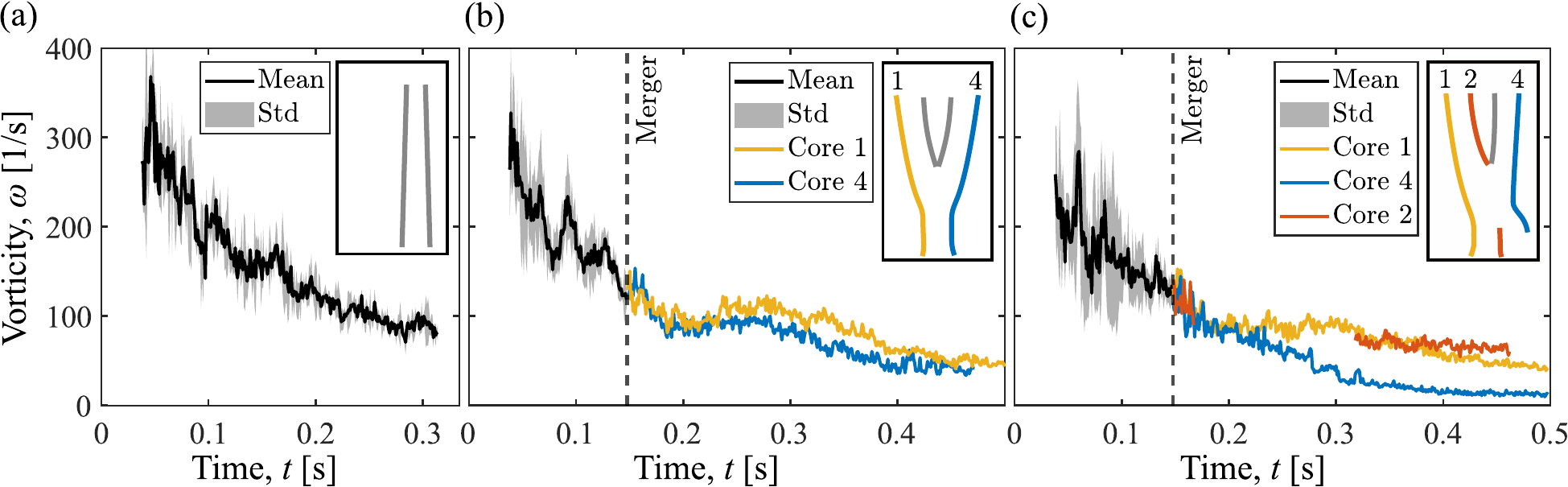}
    \caption{The local peak vorticity magnitudes as a function of time in experiments with (a) a single undisturbed ring, \textbf{(b)} two symmetrically colliding rings ($\Delta t<0.5$ ms), and \textbf{(c)} two asymmetrically colliding rings ($\Delta t=2.5$ ms). The gray areas denote the deviation of peak vorticities in identified vortex cores from the mean of their peak vorticities.}
    \label{fig:meanvort}
\end{figure*}

\subsection{\label{subsec:vel}Translational velocity}
The translational velocities of vortex rings are shown in \textbf{figure~\ref{fig:vels}a-c} for the same three model cases as in previous sections. The "initial" translational velocity $v_i$ of a ring is approximated by fitting a line to the mean depth data of vortex core pairs between 38-73 ms after the first drop impact. The "final" translational velocity $v_f$ is approximated by fitting a line to the mean depth data of vortex core pairs between 288-323 ms after the first drop impact. In \textbf{figure~\ref{fig:vels}d}, $v_i$ of the first and second ring is shown with black and red, respectively, for all experiments performed in this work. The initial velocities of the vortex rings are similar in all double and single drop experiments with the exception of regime (iii). In regime (iii), the initial velocity of the first rings is similar to other experiments, but the second ring has a noticeably lower initial velocity. The lower initial velocity in regime (iii) is assumed to be caused by the perturbed crater-vortex interaction at the creation of the second vortex ring caused by the first drop impact. Santini \textit{et al.}\cite{santini2017experimental} found in their study of synchronized double drop impacts that the translational velocity of the vortical structures generated by two close-by impacts is lower than in corresponding single drop impact cases, confirming that the crater–crater interaction affects also the vortex dynamics. Our results also show lower initial translational velocities in the synchronized and almost synchronized double drop cases compared to single rings.

In \textbf{figure~\ref{fig:vels}e}, $v_f$ of all experiments are plotted, showing that vortex ring interactions decrease the translational velocities of the rings. The final velocity of the first ring increases as $\Delta t$ between the drop impacts increases. In the unsynchronised cases in regime (ii), the translational velocity of the merged ring is lower than in the synchronised case, since the core of the second ring decays and stagnates as seen previously in \textbf{figure~\ref{fig:spatial_traj}c}. In regime (iii), the variation in the translational velocities is the largest. The second rings are significantly impaired by the first drop impact, and they are unable to penetrate deep into the pool. In regime (iv), the effects of the vortex ring interactions for the translational velocity start to disappear.

Northrup\cite{northrup1911experimental} reported in 1911 that a ring formed by the merging of two simultaneously generated parallel rings continued to travel with the same approximate velocity as a single ring, but this observation was not quantitative due to the experimental limitations. Northrup also remarked the complicated vibratory motion of merged rings in planes normal and parallel to the direction of motion of the ring. Our results indicate that the collision of vortex rings decreases the translational velocity, since the difference between initial velocities in synchronized rings and single rings is smaller than the difference between the final velocities of synchronized and single rings. However, the decrease in velocity is not very substantial and, as noted previously, the three-dimensional shape of the merged ring is deformed and continues oscillating after the collision. Our 2D visualization does not provide complete information about the location of the merged ring.


\subsection{\label{subsec:vorticity}Vorticity}
\textbf{Figure~\ref{fig:meanvort}} shows the temporal evolution of the local peak vorticities around all visible cores. The local peak vorticities were determined by taking the maximum vorticity magnitudes found within nine closest vectors around the identified vortex core centre points. \textbf{Figure~\ref{fig:meanvort}a} presents the mean of peak vorticity magnitudes of both cores of a single undisturbed ring. \textbf{Figure~\ref{fig:meanvort}b} shows the evolution of vorticity in a synchronised double drop experiment with $\Delta t<0.5$ ms. Before merging, the peak vorticities of the four visible cores are very similar to each other. The mean  magnitude of the local peak vorticities around the four cores is plotted with the shaded area indicating the standard deviation. After merging, the peak vorticity of the two visible cores is plotted separately.

In the unsynchronised case ($\Delta t =2.5$ ms) presented in \textbf{figure~\ref{fig:meanvort}c}, the vorticities of the cores start to diverge after merging. Notably, the vorticity of the remaining visible core (core 4) of the second vortex ring rapidly decreases to the level of background noise, and a third core (core 2) reappears with a similar vorticity magnitude as the first core (core 1) but with the opposite sign. The local peak vorticities confirm the observation that after the emergence of a third visible core (core 2), one of the other two (core 4) diffuses away.

The annihilation of vorticity and reconnection of vortex lines at the point of collision does not propagate along the vortex lines in a symmetric collision, since the vorticity of the surviving cores does not decrease compared to the undisturbed vortex cores. This observation is a direct interpretation of the local peak vorticities of the remaining visible vortex cores plotted as a function of time in \textbf{figure~\ref{fig:meanvort}}. The temporal evolution of the peak vorticities in the non-colliding cores matches the evolution of single undisturbed vortex rings. 

\begin{figure*}
    \centering
    \includegraphics[width=0.7\linewidth]{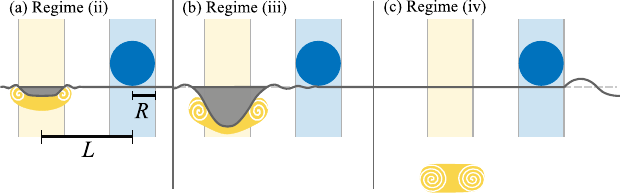}
    \caption{Illustration of capillary waves travelling from the first impact site to the second. \textbf{(a)} In regime (ii), the capillary waves of the first impact do not reach the impact site before the second crater has started forming. \textbf{(b)} In regime (iii), the capillary waves affect the impact and crater formation from the beginning. \textbf{(c)} In regime (iv), all of the waves caused by the first impact have passed through the second impact site by the time of the second impact.}
    \label{fig:res_schematic}
\end{figure*}

\subsection{\label{subsec:crater}Effect of capillary waves on crater evolution and vortex ring interactions}

The observed vortex ring interaction regimes can be explained by modelling how the capillary waves formed by the first drop impact affect the crater formation and vortex ring creation of the second drop impact, as illustrated in \textbf{figure~\ref{fig:res_schematic}}. The crater caused by a single drop impact has rotational symmetry. The craters of two synchronised drop impacts are slightly angled toward each other (\textbf{supplementary figure S3a}). Minor unsynchronisation ($0.5\leq\Delta t<7$ ms) between the two impacts in regime (ii) causes the second crater to be slightly more angled toward the first (\textbf{supplementary figure S3b}). Critically unsynchronised impacts ($7\leq\Delta t<80$ ms) in regime (iii) lead to a significantly distorted second crater (\textbf{supplementary figure S3c}). With major unsynchronisation ($\Delta t\geq 80$ ms) in regime (iv), the shapes of both craters are undisturbed (\textbf{supplementary figure S3d}).


The initial velocities of the second rings in regime (iii) in \textbf{figure~\ref{fig:vels}d} indicate that these are impaired from the moment of detaching from the crater, meaning that the generation of the second ring is affected by the surface effects caused by the first impact. By following the progression of the capillary waves on the surface of the pool liquid in the experiments (see \textbf{supplementary movies M3-M4}), the transition from regime (ii) to regime (iii) seems to occur when the capillary waves created by the first impact have reached the impact site of the second drop by the time the second crater starts to form. Also, the transition from regime (iii) to regime (iv) happens when the last wave of the first crater has already passed the second impact site by the time of the second impact. This observation allows us to generalize our results to other situations with different inter-droplet distances $L$, surface tensions $\gamma$, fluid densities $\rho$, droplet sizes $R$ and impact velocities $v$. We can estimate the critical time differences for the regime transitions by calculating capillary wave velocities and the time it takes for the crater to dissipate past the second impact site.

The phase velocity of capillary waves is given by 

\begin{equation}
    v_\mathrm{c} = \sqrt{\frac{2\pi \gamma}{\rho \lambda}},
\end{equation}
where $\lambda$ is the wavelength \citep{gennes2004capillarity}. By using density $\rho=1000$ kg/m$^3$, the observed wavelength of $\lambda=2$ mm of the first visible capillary waves (\textbf{supplementary figure S4a}), and the measured surface tension $\gamma=0.062$ N/m, we arrive at a phase velocity of $v_\mathrm{c, frist}=0.44$ m/s. The observed velocity of the first visible capillary wave troughs was $v_\mathrm{frist}=0.46\pm0.01$ m/s. With the observed velocity, capillary waves reach the middle of the second impact site $t_\mathrm{c}=L/v_\mathrm{frist}=13$ ms after the first impact. Since the generated vortex ring detaches from the crater approximately $t=15 \text{ ms}>t_\mathrm{c}$ after the impact, we can assume that capillary waves affect the crater evolution even in the case of synchronised drop impacts. However, when the impacts are simultaneous, both craters should be affected by the capillary waves equally. The recording from the synchronized experiment ($\Delta t<0.5$ ms), supports this, as the craters seem to be equally tilted toward each other and evolve in-synch. Therefore, the difference between regimes (i) and (ii) is assumed to be caused by the increasing asymmetry of the evolving craters in regime (ii): the first vortex ring may be able to detach from the crater before the capillary waves from the second impact reach the first crater, whereas the second crater is affected by the capillary waves of the first impact before the second ring detaches from it.

The transition between regimes (ii) and (iii) is observed to occur when the capillary waves from the first drop interfere with the impact region of the second drop by the time the second crater starts to form. The distance from the centre of the first impact site to the closest point of the second impact site is $L-R$, as illustrated in \textbf{figure~\ref{fig:res_schematic}}. This transition time $\tau_{\mathrm{(ii)}\rightarrow\mathrm{(iii)}}$ can be written as

\begin{equation}
    \tau_{\mathrm{(ii)}\rightarrow\mathrm{(iii)}}= \frac{L-R}{v_{\mathrm{first}}}-t_\mathrm{cr},
    \label{transiitoiii}
\end{equation}
where $v_{\mathrm{first}}$ is the velocity of the first capillary waves and $t_{\mathrm{cr}}$ the time it takes for the crater to start forming after impact. Based on our observations, it takes 9.8 ms for the capillary waves to reach the closest point of the second impact site and $t_\mathrm{cr}=3.5$ ms. This gives a theoretical estimate of $\tau_{\mathrm{(ii)}\rightarrow\mathrm{(iii)}} = 6.3$ ms for our experiments, defining when the capillary waves affect the second crater formation from the beginning. This matches well with the observed regimes in the experiments, where the transition from regime (ii) to regime (iii) happens at $\Delta t=5.3\pm1.8$ ms (\textbf{figure~\ref{fig:angle}d-e} and \textbf{figure~\ref{fig:vels}d-e}). With the $\Delta t = 6.3$ ms time difference, the capillary waves of the second impact no longer reach the first crater before the detachment of the first vortex ring, since it takes $t=6.3+9.8=16.1$ ms for the waves caused by the second impact to reach the closest point of the first impact site and the first ring detaches at $t=15$ ms after the first impact. This explains the observed change in behaviour of the vortex rings in regime (iii) compared to regime (ii). \textbf{Figure~\ref{fig:angle}d} showed that in regime (ii), both rings tilted, but unequally, which could be explained by both crater evolutions being affected by capillary waves to some extent but unequally. In regime (iii), however, we saw that the first rings moved straight down like undisturbed rings, but the second rings tilted. Likewise, in regime (iii) in \textbf{figure~\ref{fig:vels}d} the initial translational velocity of the first ring was unaffected, but the initial translational velocity of the second ring was significantly reduced. This behaviour matches the first crater evolution not being affected by the second impact, but the second crater evolution being affected from the beginning by the waves caused by the first impact. 

Regime (iii) contains the most diverse set of behaviours (see \textbf{supplementary movies M4, M6, M7} and \textbf{M8}). Common features are the defining impairment of the second ring, nearly undisturbed first rings, and distorted second craters. However, there are many ways the second crater can be distorted and the second ring has varying levels of impairment -- in one experiment the second ring broke into two weak rings as it detached from the crater (see \textbf{supplementary movie M7}). The reason for the large differences lies in the fact that the second drop impacts at different phases of the capillary waves and dissipating first crater. In the experiments, the crater starts forming at $t_{\mathrm{cr}}=3.5$ ms, a maximum crater depth is seen at $t_{\mathrm{cr, max}}=10$ ms, and the generated vortex ring detaches from the crater approximately 15 ms after the impact. 
The last visible trough of the retracting crater travels at velocity $v_{\mathrm{last}}=0.20\pm0.01$ m/s. The impact sites are back at rest approximately 90 ms after the impact, which should mark the transition from regime (iii) to regime (iv). 
The transition time difference $\tau_{\mathrm{(iii)}\rightarrow\mathrm{(iv)}}$ can be expressed as

\begin{equation}
    \tau_{\mathrm{(iii)}\rightarrow\mathrm{(iv)}}= \frac{L+R}{v_{\mathrm{last}}}+t_{\mathrm{d}},
    \label{transiiitoiv}
\end{equation}
where $v_{\mathrm{last}}$ is the velocity at which the final waves of the dissipating crater travel and $t_{\mathrm{d}}$ the time it takes for these final waves to become visible after the drop impact. In our case ($t_{\mathrm{d}}=20\pm5$ ms), this theoretical transition time is $\tau_{\mathrm{(iii)}\rightarrow\mathrm{(iv)}}=57.5\pm$5.4 ms. In the experiments, the transition was observed to occur when the time difference between the two drop impacts was $56.5<\Delta t <80$ ms (\textbf{figure~\ref{fig:angle}d-e} and \textbf{figure~\ref{fig:vels}d-e}). 

All observed and theoretical transition times are summarised in \textbf{table~\ref{tab:transitiontimes}} and are in very good agreement with each other. The time taken for the liquid surface to calm down after the first drop impact can thus be used to estimate when the second nearby drop can impact the pool without having its subsequent vortex ring impaired by the impact of the first drop.

\begin{table}
\caption{\label{tab:transitiontimes} Transition times between the different vortex ring interaction regimes (i-iv) based on observations and theory.}
\begin{center}
\begin{tabular}{c|c|c}
\hline
\hline
        Transition & $\tau_{\mathrm{obs}}$ (ms) & $\tau_{\mathrm{theory}}$ (ms)  \\ \hline
        (i)-(ii) & $\leq$0.5 & $>0$ \\ 
        (ii)-(iii) & $5.3\pm1.8$ & 6.3 $\pm$ 1.1 \\ 
        (iii)-(iv) & $68\pm12$ & 57.5 $\pm$ 5.4 \\
        \hline
\hline
    \end{tabular}
\end{center}
\end{table}

\section{\label{sec:conclusion}Conclusion}
In this work we have investigated the evolution and interaction of two vortex rings generated by drop impacts with a varying time difference and compared them to single drop-formed vortex rings. Using particle image velocimetry, we track the locations of the vortex cores to study their trajectories, as well as the evolution of angles, translational velocities, peak vorticities, and distances between the cores. We observe four different vortex ring interaction categories which depend on the time difference between the two impacting droplets. At $\Delta t<0.5$ ms, the generated vortex rings collide and merge symmetrically. Larger time differences drastically affect the collision and merging, which either becomes asymmetric and incomplete ($0.5 < \Delta t<7$ ms) or does not happen at all ($\Delta t \geq 7$ ms). At $7 \leq \Delta t<80$ ms, the second vortex ring creation is heavily impaired due to the capillary waves created by the first drop impact. At $\Delta t\geq 80$ ms, the first impact no longer hinders the formation of the second ring and eventually the time difference between the impacts is large enough for them to be effectively single drop impacts leading to undisturbed vortex rings. We develop a model based on the velocity of the capillary waves created by the impacting droplets to describe the transition times between the different regimes. The goal of this study was to probe the effect of $\Delta t$ on the above mentioned vortex ring properties. Future work could focus on expanding the parameter space to cover a wider range of initial conditions (e.g., drop size, impact velocity, Weber number, and inter-droplet distance) and perform more advanced analysis of the flow (e.g., circulation, enstrophy, vortex momentum, and viscous diffusion) either through new experiments or numerical simulations. Applying more evolved flow visualizations to capture the complete 3D flow could also provide interesting information about the 3D aspects of the vortex ring shapes. Finally, our experimental results can also be used in future theoretical work developing a more advanced physical model of the observed phenomenon. Our work provides insight for future drop-induced mixing applications, where the spatial and temporal distance between the impacting drops needs to be adjusted to ensure optimal vortex ring generation and an undisturbed evolution of the rings deep into the liquid pool.









\section*{Supplementary data.}{\label{SupMat}Supplementary material and movies are available at https://doi.org/..}

\begin{acknowledgments}
This work was funded by the European Union (ERC, SWARM, 101115076; M.B.) and the Research Council of Finland Fellowship (MesoSwim, 354904; M.B.). Views and opinions expressed are however those of the author(s) only and do not necessarily reflect those of the European Union or the European Research Council. Neither the European Union nor the granting authority can be held responsible for them.

\end{acknowledgments}
\section*{Author declarations}
The authors have no conflicts of interest to disclose. A.H.: Conceptualization; investigation; data curation; formal analysis; software; visualization; writing – original draft (lead); writing – review and editing (equal). G.M.B.: Conceptualization; methodology; formal analysis (supporting); software; writing – review and editing (equal). M.B.: Conceptualization; funding acquisition; project administration; supervision; visualization; resources; writing – original draft (supporting); writing – review and editing (equal).

\section*{Data Availability Statement}
The data that support the findings of this study are openly available in Zenodo at http://doi.org/10.5281/zenodo.20624416, reference number \cite{HuttunenZenodo26}. 

\section*{References}
\bibliography{aipsamp}

\end{document}